\documentclass[slac_one]{revtex4}
\usepackage{graphicx}
\usepackage{subfigure}
\usepackage{fancyhdr}
\usepackage{xspace}
\usepackage{amssymb}
\begin{document}

\title{Radiative Penguin decays at Belle}

\author{Jin Li (for the Belle Collaboration) }
\affiliation{University of Hawaii, Honolulu, HI 96822, USA}

\begin{abstract}
We present recent progresses in radiative penguin
decays of $B$ meson using a large sample of $B\bar{B}$ pairs recorded at
the $\Upsilon(4S)$ resonance with the Belle detector at the KEKB
asymmetric energy $e^+e^-$ collider.  We report precise measurement
of inclusive $b\to s\gamma$ branching ratio with cut $E_\gamma > 1.7$ GeV,
first measurement of time-dependent CP-violation in $B^0\to K_s\rho^0\gamma$,
measurement of $B^+\to K^+\eta'\gamma$ branching fraction, and improved
branching fraction results for $B^0\to(\rho,\omega)\gamma$ with new $CP$
and isospin violation results in the mode.

\end{abstract}
\maketitle
\thispagestyle{fancy}

\section{Introduction}
Radiative $B$ decay proceeds through a penguin loop diagram in the
Standard Model (SM).  New particles beyond the SM may contribute in
the loop diagrams.  In this paper, we present recent Belle results on
various radiative $B$ decay topics below.  Inclusive $b\to s\gamma$
measurements will constrain New Physics from the branching fraction as
well as determining $b$-quark mass and motion from photon energy
spectrum. Time dependent CP asymmetry in $b\to s\gamma$ transition can
probe for right-handed coupling which is not present in SM.
Measurements of exclusive mode $B\to \eta'K\gamma$ and
$B\to(\rho,\omega)\gamma$ can probe deviations from SM by the
branching fraction and $CP$ asymmetry.

In all these studies, large continuum background from $e^+e^-\to q\bar
q$ events ($q = u,d,s,c$) has been suppressed using event topology
variables.  In exclusive modes, two kinematic variable $M_{bc} =
\sqrt{E^2_\mathrm{beam} - p^{*2}_B}$ and $\Delta E=E_B^* -
E_\mathrm{beam}$ are commonly used.  Here $E_\mathrm{beam}$ is the
beam energy and $E^*_B$ and $p^*_B$ is the energy and momentum of a
$B$ candidate, all defined in the $\Upsilon(4S)$ frame.

\section{Inclusive $b\to s\gamma$ measurement}
In this fully inclusive method, $b\to s\gamma$ decay is studied using
a $604.6\mathrm{fb}^{-1}$ ON data sample taken at the $\Upsilon(4S)$
resonance, and $68.3\mathrm{fb}^{-1}$ OFF data sample taken at an
energy $60\,\mathrm{MeV}$ below the resonance.  Hard photon is
selected from well isolated ECL clusters with the shower
shape consistent with a photon, and required to have a center-of-mass energy
$E^*_\gamma >1.4$ GeV.  $\pi^0$ and $\eta$ vetoes are then applied.

\begin{figure}\centering
\subfigure[]{\includegraphics[width=0.3\textwidth]{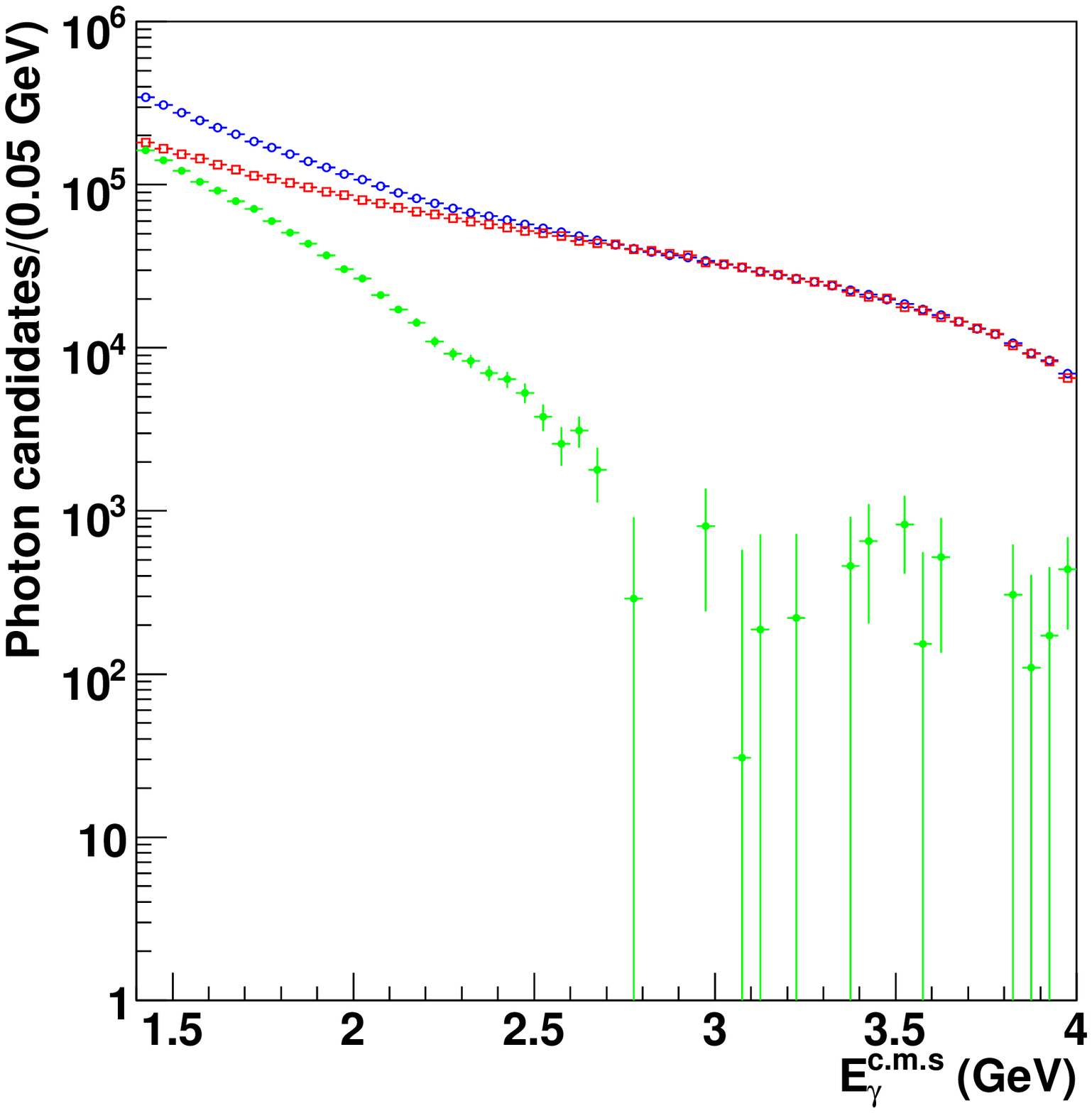}}%
\subfigure[]{\includegraphics[width=0.3\textwidth]{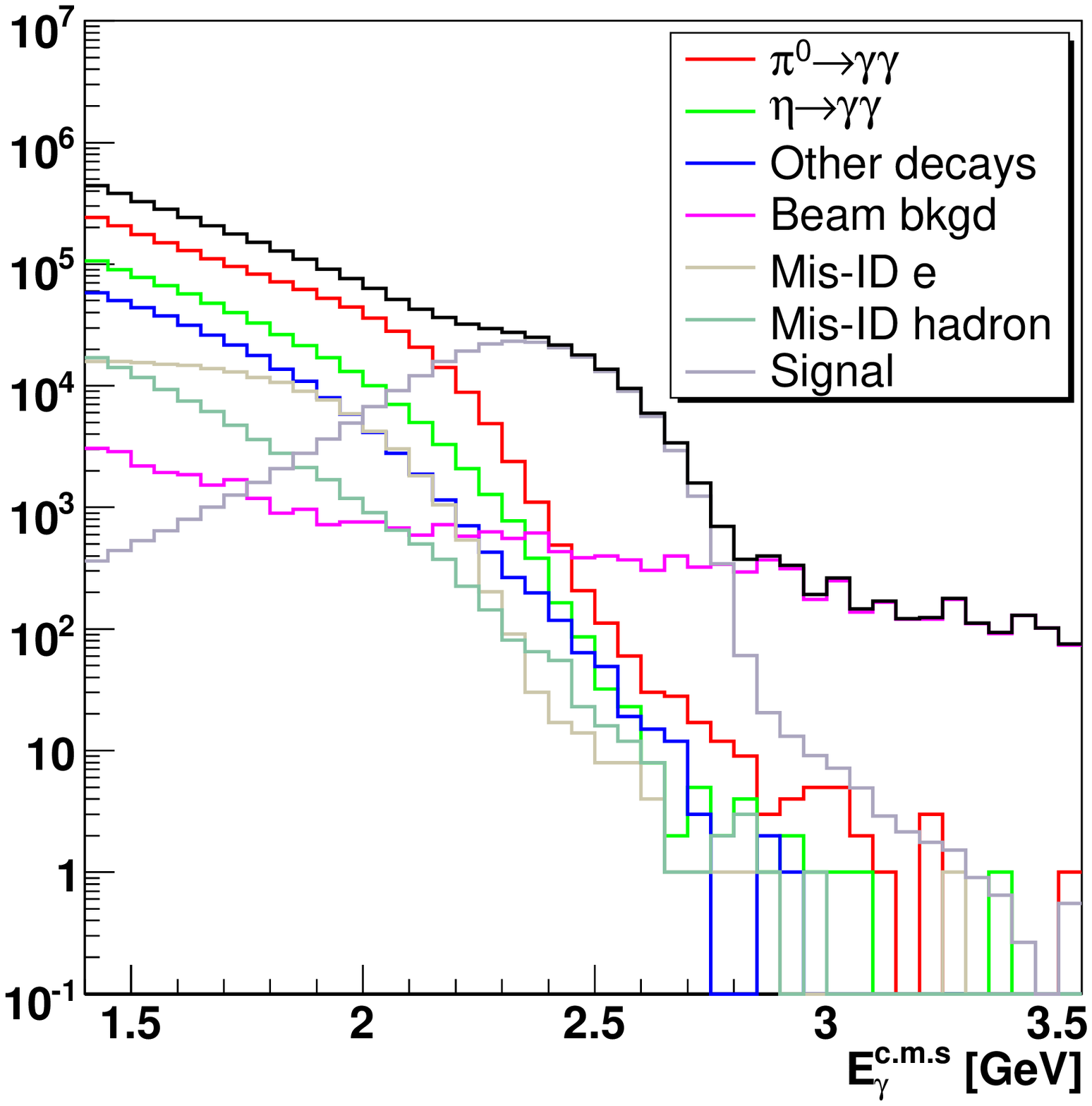}} %
\subfigure[]{\includegraphics[width=0.3\textwidth]{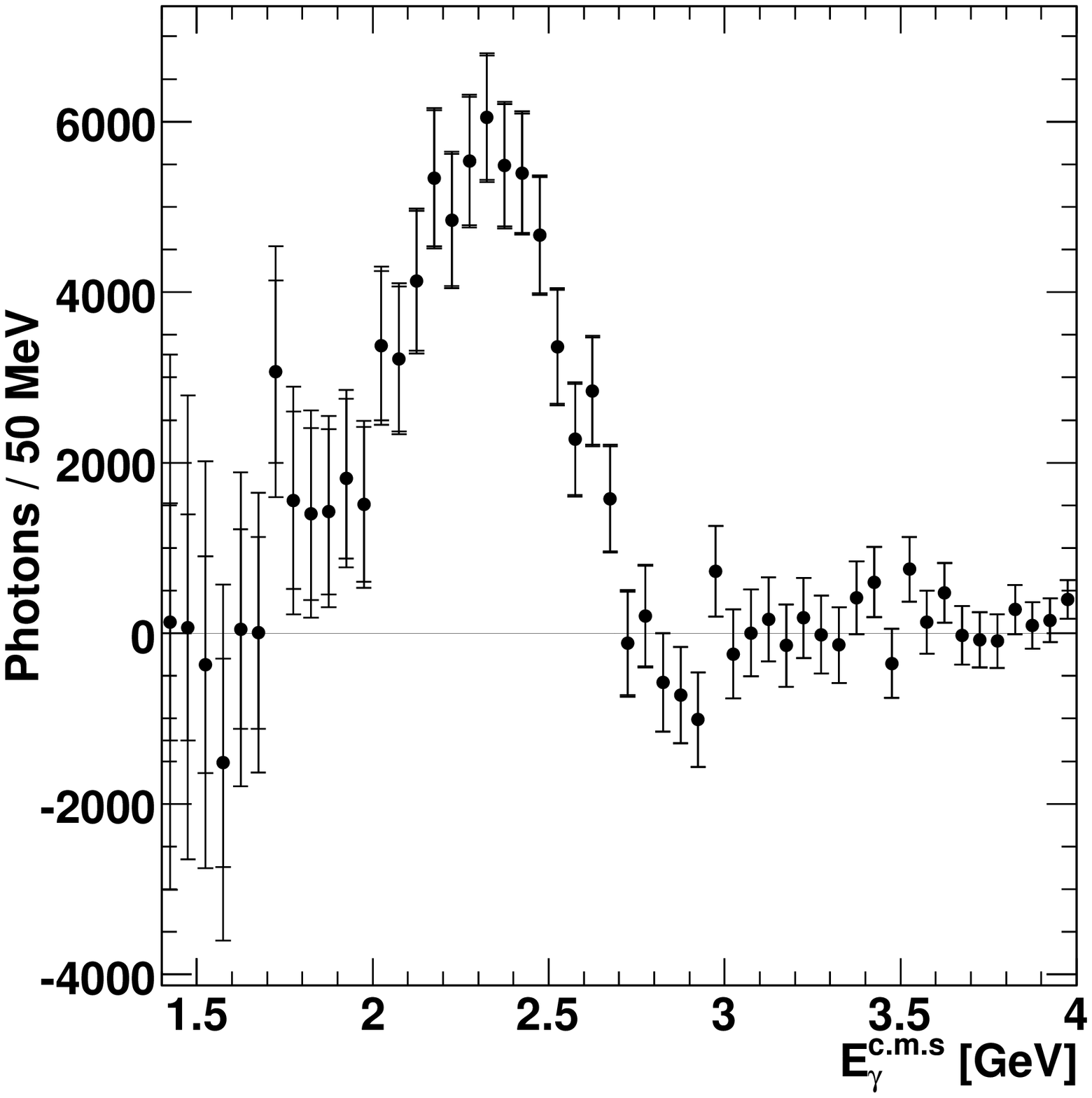}}
\caption{\label{fig:Bdata}
(a) ON data (open circle), scaled OFF data (open square) and
continuum background subtracted (filled circle) photon energy
spectra of candidates in the c.m.s frame. (b)
The spectra of photons from $B$-meson decays passing
selection criteria as predicted using a MC sample. 
(c)  The extracted photon energy spectrum of $B\to X_{s,d}\gamma$. 
 The two error bars show the statistical and total errors.
}
\end{figure}

First, the photon energy spectrum measured in OFF data is scaled by
luminosity to the expected number of non-$B\bar B$ events in ON data
and subtracted, shown in Fig.~\ref{fig:Bdata}(a).  Here the OFF
spectrum has been corrected due to the slight lower energy than ON
events.  Then the backgrounds coming from non-primary photons from $B$
mesons subtracted from the obtained spectrum.  From Monte Carlo (MC)
study, six background categories are considered
(Fig.~\ref{fig:Bdata}(b)): (i) photons from $\pi^0\to\gamma\gamma$;
(ii) photons from $\eta\to\gamma\gamma$; (iii) other real photons;
(iv) ECL clusters not due to single photons (mainly $K^0_L$'s and
$\bar{n}$'s); (v) Electrons misidentified as photons; and (vi) beam
background.  For each of the background category, the shape and yield
is corrected from MC using data by selecting them.  Then, for each
selection criterion used in this analysis, the data-MC efficiency
ratio is obtained using appropriate control samples, and is then used
to scale the MC background sample.  For example, the $\pi^0$ veto
efficiency is studied using $D^{*+}\to D^0(K^-\pi^+\pi^0)\pi^+$ decays
where only one $\gamma$ from $\pi^0$ is reconstructed.

The raw spectrum shown in Fig.~\ref{fig:Bdata}(c) is obtained after
subtracting the six background categories after scaling.  The raw
spectrum is then corrected by three step procedure: (i) divide by the
efficiency of the selection; (ii) perform an unfolding which removes
the distortion by caused ECL; (iii) divide by efficiency of detection.
Then, two additional corrections are applied: (i) to remove cabibbo
suppressed $B\to X_d\gamma$ decays, using the ratio of branching
fractions; (ii) to correct the measurements to $B$-meson rest frame
from $\Upsilon(4S)$ frame (boost correction).  The partial
branching fraction, first moment (mean) and second central moment
(variance) of the photon energy spectrum from $B$ decay are obtained
as:~\cite{Abe:2008sx}
$\mathcal{B}\left( B\to X_s \gamma \right) = \left( 3.31 \pm 0.19
  \pm 0.37 \pm 0.01 \right )\times10^{-4}$;
$\left< E_\gamma \right> = 2.281 \pm 0.032 \pm 0.053 \pm 0.002\,\mathrm{GeV}$;
$\left <E_\gamma^2\right>-\left<E_\gamma\right>^2 =
 0.0396 \pm 0.0156 \pm 0.0214 \pm 0.0012\,\mathrm{GeV}^2$,
where the errors are statistical, systematic and from the boost
correction, respectively.

\section{Time-dependent $CP$ Asymmetries in $B^0\to K^0_S\rho^0\gamma$ Decay}
In this mode, signal $B^0$ decay vertex can be reconstructed from two
charged pions to calculate the decay time difference $\Delta t$
between signal and tag side $B$, thus avoiding $K_S^0$ vertexing.  The
$B^0\to K_S^0\rho^0 \gamma$ candidates are selected from the
$K_S^0\pi^+\pi^- \gamma$ sample by requiring the $\pi^+\pi^-$
invariant mass to lie in the $\rho^0$ region,
$0.6\,\mathrm{GeV}/c^2<m_{\pi\pi}<0.9\,\mathrm{GeV}/c^2$.  We first
measure the effective $CP$-violating parameters,
$\mathcal{S}_{\rm{eff}}$ and $\mathcal{A}_{\rm{eff}}$, using the final
sample and then convert them to the $CP$-violating parameters of
$B^0\to K_S^0 \rho^0\gamma$ using a dilution factor $\mathcal{D}$.

\begin{figure}
\includegraphics[width=0.4\columnwidth]{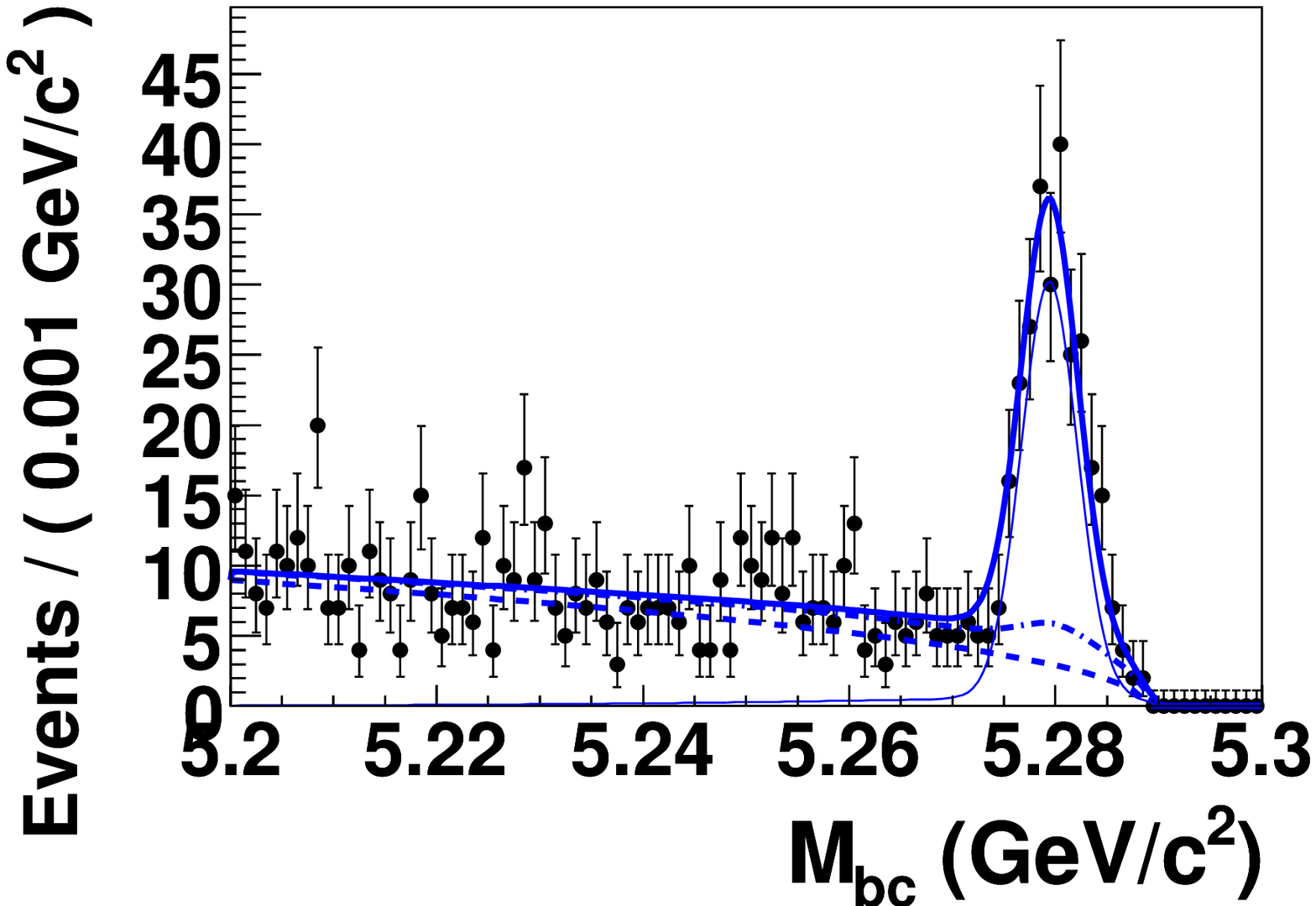}\hspace{20pt}
\includegraphics[width=0.3\columnwidth]{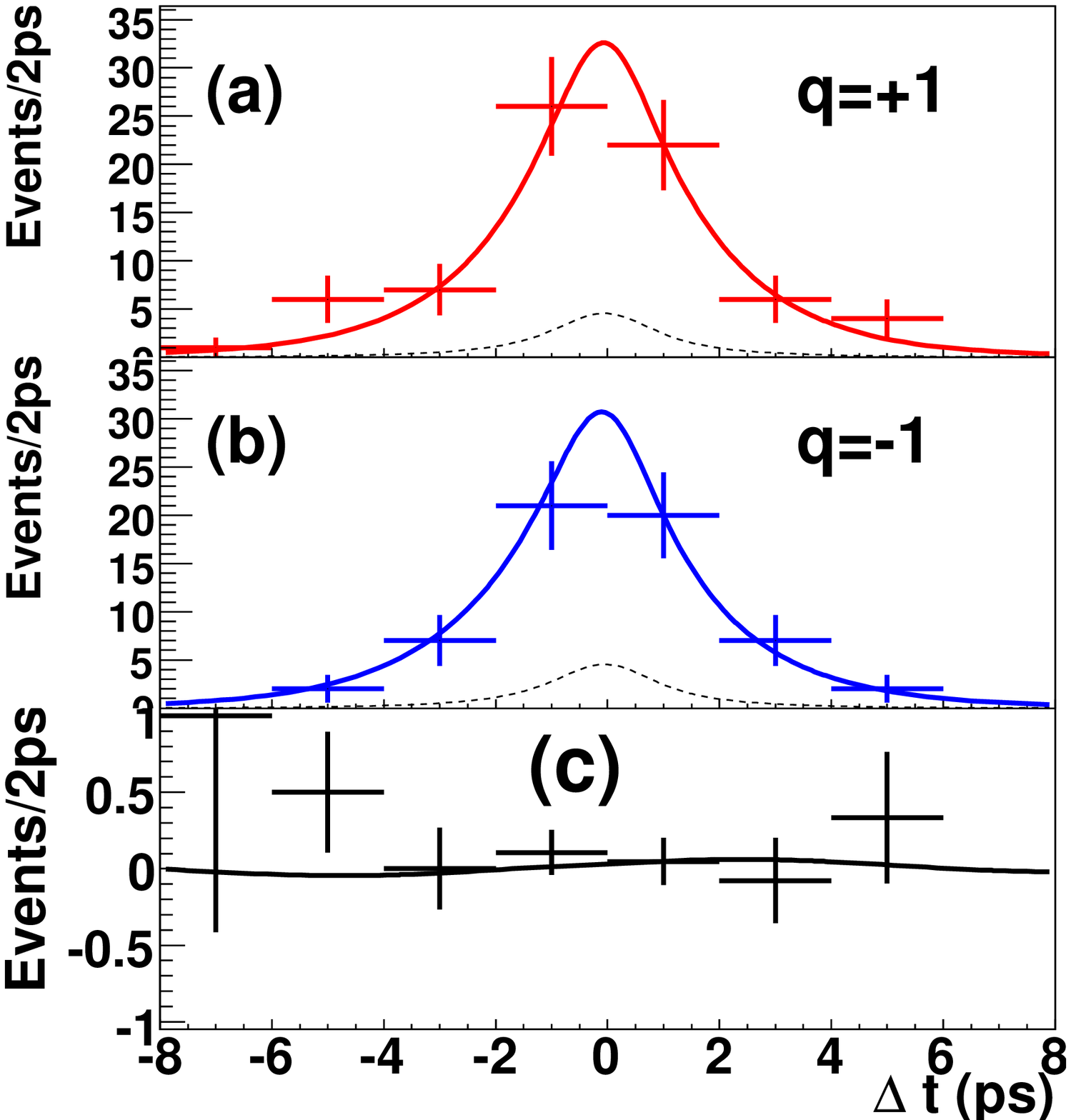}
\caption{\label{fig:ksrhog}
(Left) $M_\mathrm{bc}$ distributions for 
 $B^0\to K_S^0\pi^+\pi^-\gamma$ events in $\rho^0$ region.
The dashed and dash-dotted curves are the $q\bar q$ and all BG.
The thin curve is the total signal and the thick curve
is the total PDF. (Right) Fit projections on the $\Delta t$ distributions
for events good-tagged as (a) $B^0$ and (b) $\overline{B}^0$.
The raw asymmetry as a function of $\Delta t$ is shown in (c).
}
\end{figure}
Fig.~\ref{fig:ksrhog}(left) shows a fit to $M_\mathrm{bc}$ in $\Delta
E$ signal region after vertexing.  We obtain $212\pm 17$ total signal
yield from the total 299 events in the signal $M_\mathrm{bc}$ region.
The $\mathcal{S}_\mathrm{eff}$ and $\mathcal{A}_\mathrm{eff}$
parameters are then extracted from an unbinned maximum likelihood fit
to the $\Delta t$ distribution, as $\mathcal{S}_\mathrm{eff} = 0.09\pm
0.27(\mathrm{stat.})  ^{+0.04}_{-0.07}(\mathrm{syst.})$ and
$\mathcal{A}_\mathrm{eff} = 0.05\pm 0.18(\mathrm{stat.})  \pm
0.06(\mathrm{syst.})$.  The $\Delta t$ projection for the fit is shown
in Fig.~\ref{fig:ksrhog}(right).

The parameter $\mathcal{S}_\mathrm{eff}$
is related to $\mathcal{S}$ for $K_S^0\rho^0\gamma$
with a dilution factor $\mathcal{D}$:
\[
\mathcal{D} = \frac{\mathcal{S}_\mathrm{eff}}
{\mathcal{S}_{K_S^0\rho^0\gamma}} = \frac{\int [|F_A|^2+
2\Re(F_A^*F_B) + F_B^*(\bar K)F_B(K) ]} {\int \left[|F_A|^2+
2\Re(F_A^*F_B) + |F_B|^2\right]},
\]
where $F_A,F_B$ are photon-helicity averaged amplitudes for $B^0\to
K_S^0\rho^0(\pi^+\pi^-)\gamma$ and $B^0\to
K^{*\pm}(K_S^0\pi^\pm)\pi^\mp\gamma$, respectively. From the study to
($K^+\pi^-\pi^+$) system in charged mode $B^+\to K^+\pi^-\pi^+\gamma$
using isospin symmetry, we obtain $\mathcal{D}=0.83^{+0.19}_{-0.03}$.
In summary, we measure the $CP$-violating parameter
$\mathcal{S}_{K_S^0\rho^0\gamma}= 0.11\pm
0.33(\mathrm{stat.})^{+0.05}_{-0.09}(\mathrm{syst.})$~\cite{Li:2008qm}.
This agrees with SM prediction $\mathcal{S}_{K_S^0\rho^0\gamma}\approx
0.03$~\cite{Atwood:2004jj}.

\section{Evidence for $B\to \eta'K\gamma$ Decays at Belle }
The exclusive mode $B\to \eta'K\gamma$ is analyzed using
$604.6\mathrm{fb}^{-1}$ of data, for both charged and neutral mode.
$\eta'$ mesons are reconstructed as $\eta' \to \eta \pi^{+} \pi^{-}$
and $\eta'\to \rho^0\gamma$.  $\eta$ mesons are reconstructed as
$\eta\to \gamma\gamma$ and $\eta\to \pi^+\pi^-\pi^0$. 
$D^0$ mass veto on $m_{K^+\pi^-}$ for any changed pion in the event
and $J/\psi$ mass veto on $m_{\eta'\gamma}$ are applied to suppress
$b\to c$ events.

\begin{figure}
\centering
\subfigure[]{\includegraphics[width=0.25\columnwidth]{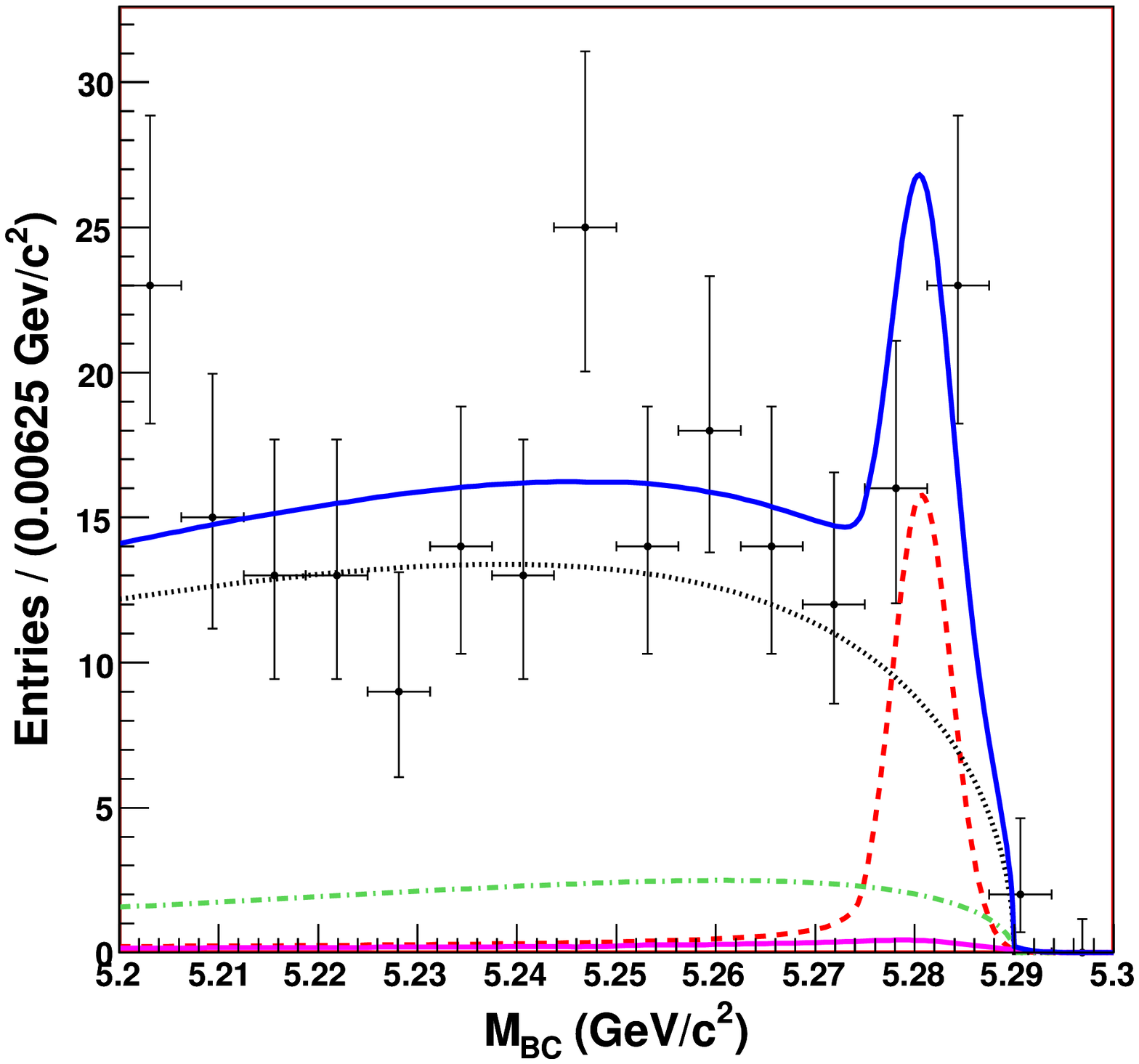}}%
\subfigure[]{\includegraphics[width=0.25\columnwidth]{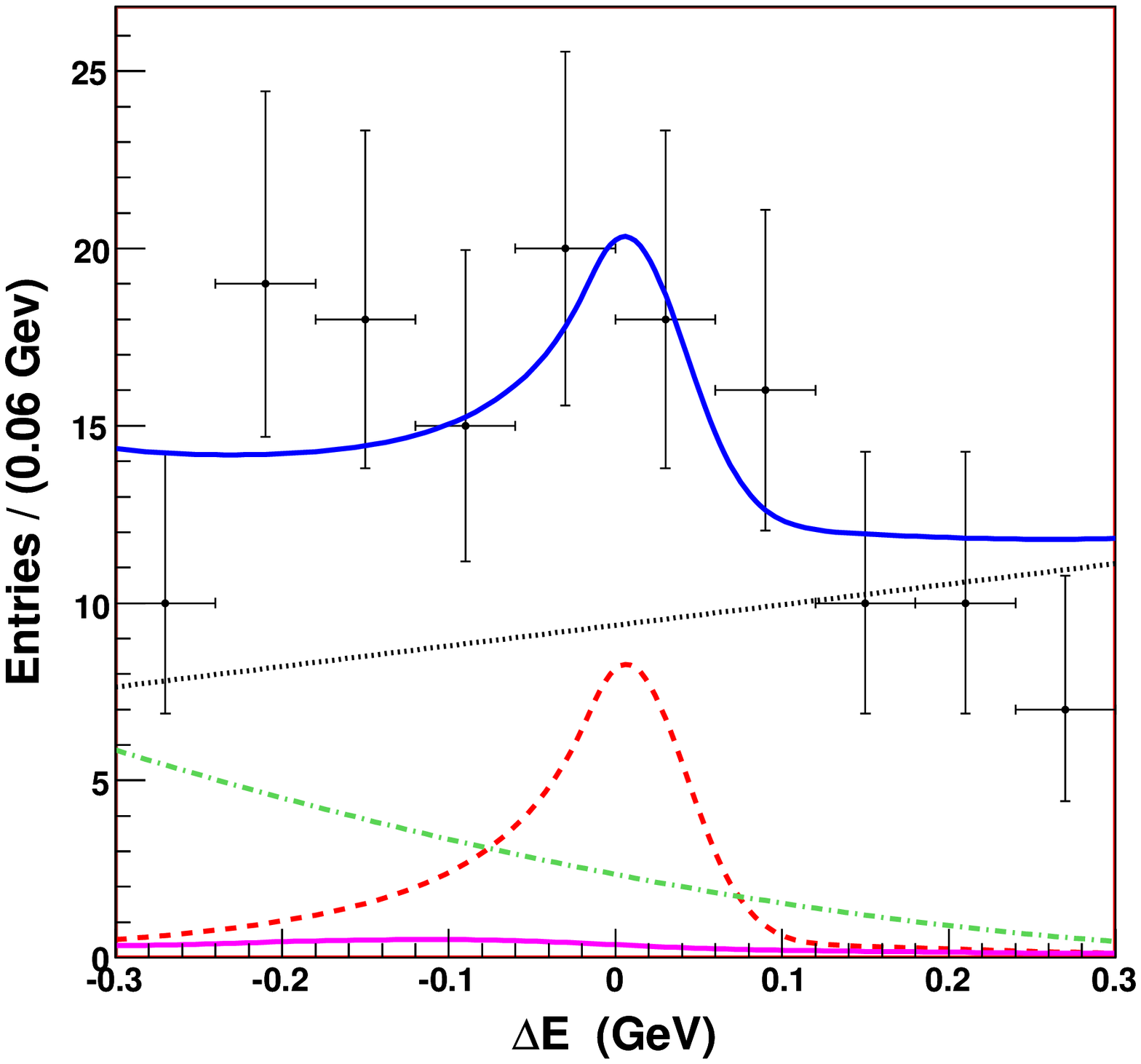}}%
\subfigure[]{\includegraphics[width=0.25\columnwidth]{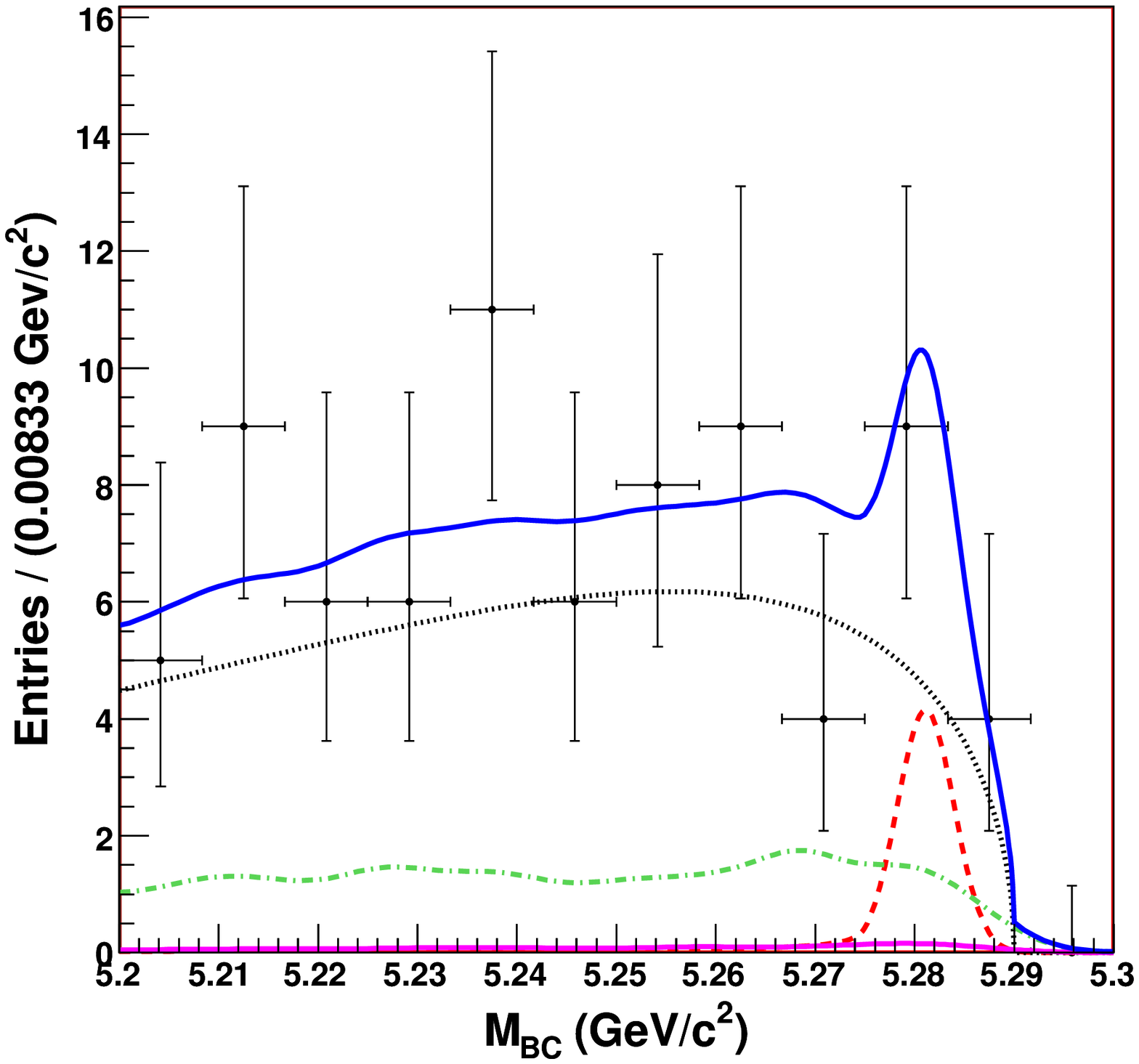}}%
\subfigure[]{\includegraphics[width=0.25\columnwidth]{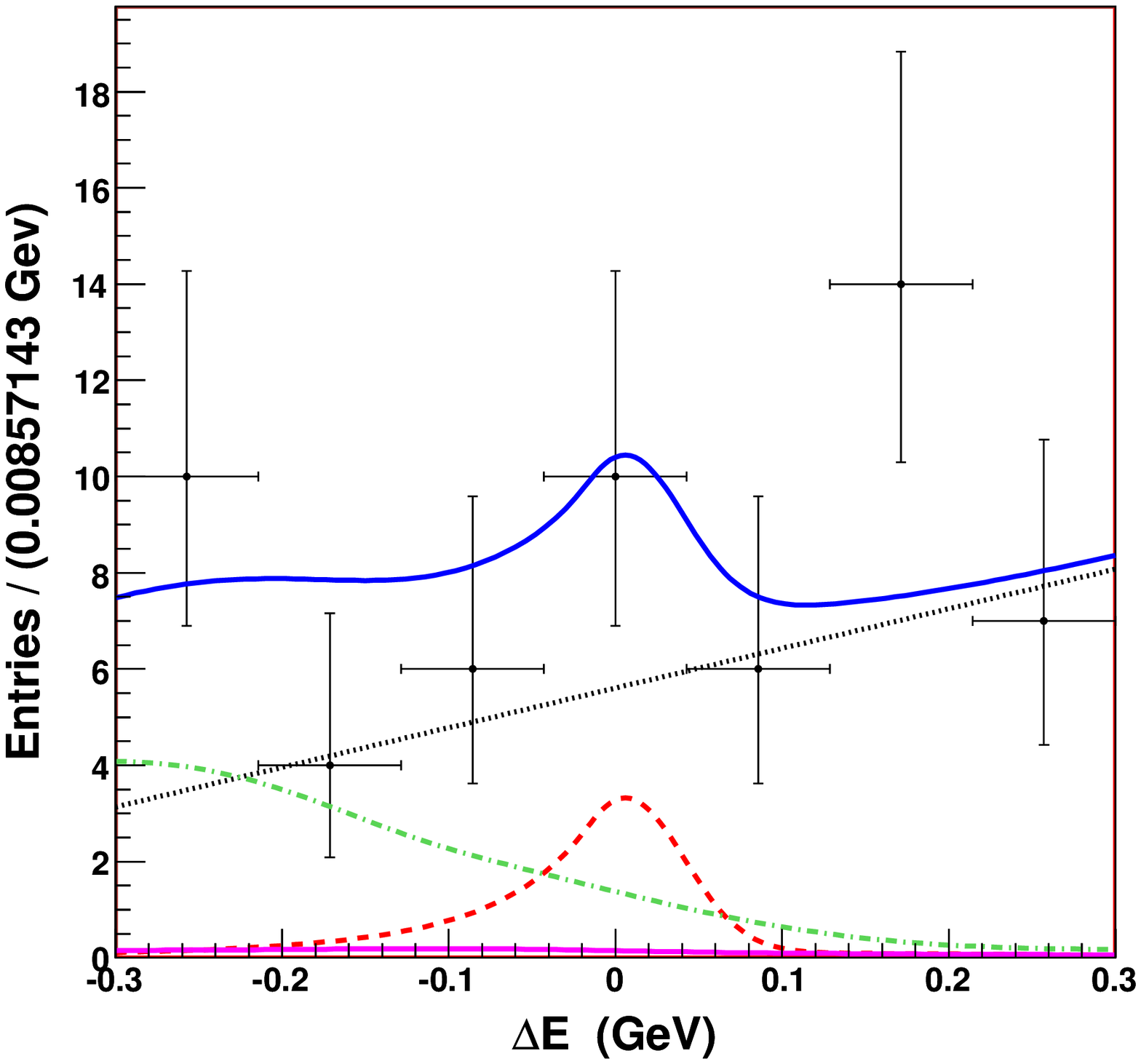}}
\vspace{-1.5\baselineskip}
\caption{Projections from the 2D fit to data. (a) and (b) plots show
$M_\mathrm{bc}$ distribution in the signal region of $\Delta E$ and
$\Delta E$ distribution in the signal region of $M_\mathrm{bc}$, for
the $B^+\to K^+\eta'\gamma$ mode.  The $K \eta' \gamma$ function is
shown in dashed red, $q\bar q$ in dotted black, $b\to c$ in
dash-dotted green, $b\to u,d,s$ in solid magenta, and the combined
function in solid blue. (c) and (d) are the same for $B^0\to
K_S^0\eta'\gamma$ mode.
\label{fig:ketapg} 
}
\end{figure}

The signal yield is extracted by fitting 2D $M_\mathrm{bc}$ and
$\Delta E$ distributions shown in Fig.~\ref{fig:ketapg}.  All final
states are combined in to one distribution for fitting for the charged
mode, so does the neutral mode.  The signal shape is calibrated using
large samples of $B\to K^*(892)\gamma$ data and MC.  The continuum
parameters are floated in the fit. 
Table~\ref{tab:ketapg} shows the fit results and branching
fractions (BFs).
The efficiencies are weighted by background-subtracted
$M(K\eta')$ data distribution.
Systematic errors are included in
the likelihood by convolving the likelihood functions to calculate
the significances and upper limits.

\newcommand{\eps}{\ensuremath{{\varepsilon}}\xspace}
\newcommand{\ul}{\ensuremath{{\cal UL}}\xspace}
\newcommand{\BToKetapgch}{\ensuremath{B^{+} \to K^{+} \eta' \gamma}\xspace}  
\newcommand{\BToKetapgn}{\ensuremath{B^{0} \to K^{0} \eta' \gamma}\xspace}  
\newcommand{\BF}{\ensuremath{{\cal BF}}\xspace}

\begin{table}
\begin{center}
\begin{tabular}{ c | c  c  c  c  c  c }\hline
{Mode} &  Yield(events) &\eps   & $\prod$ & BF($10^{-6}$) &  S'($\sigma$)  &
UL($10^{-6}$) \\\hline
\multicolumn{1}{ c |}{ \BToKetapgch }  &  $32.6^{+11.8}_{-10.8}$& 0.027 & 0.571   & $3.2^{+1.2 +0.3}_{-1.1 -0.3}$ & 3.3 & - \\ 
\multicolumn{1}{ c |}{ \BToKetapgn }   &  $5.1^{+5.0}_{-4.0}$   & 0.016 & 0.197   & $2.4^{+2.4 +0.4}_{-1.9 -0.5}$ & 1.3 & 6.3 \\\hline 
\end{tabular}
\caption{The yields, efficiencies(\eps), daughter BFs ($\prod$),
measured BFs, fit significances including systematics (S')
and upper limits (UL) for the measured decays.\label{tab:ketapg}}
\vspace{-0.5\baselineskip}
\end{center}
\end{table}

\section{$b\to d\gamma$ update by Belle}
Belle updated the measurements of $B\to\rho\gamma$, $B\to\omega\gamma$
decays using a sample of 657 million $B$ mesons.  Three signal modes,
$B^+\to \rho^+\gamma$, $B^0\to \rho^0\gamma$ and $B^0\to \omega\gamma$
are reconstructed with subdecay modes $\rho^+\to\pi^+\pi^0$,
$\rho^0\to\pi^+\pi^-$, $\omega\to\pi^+\pi^-\pi^0$. 
A helicity selection of the $\rho$ and $\omega$ is
applied to suppress decays with $\pi^0/\eta$ from $B$.

\begin{figure}\centering
\includegraphics[width=0.33\textwidth]{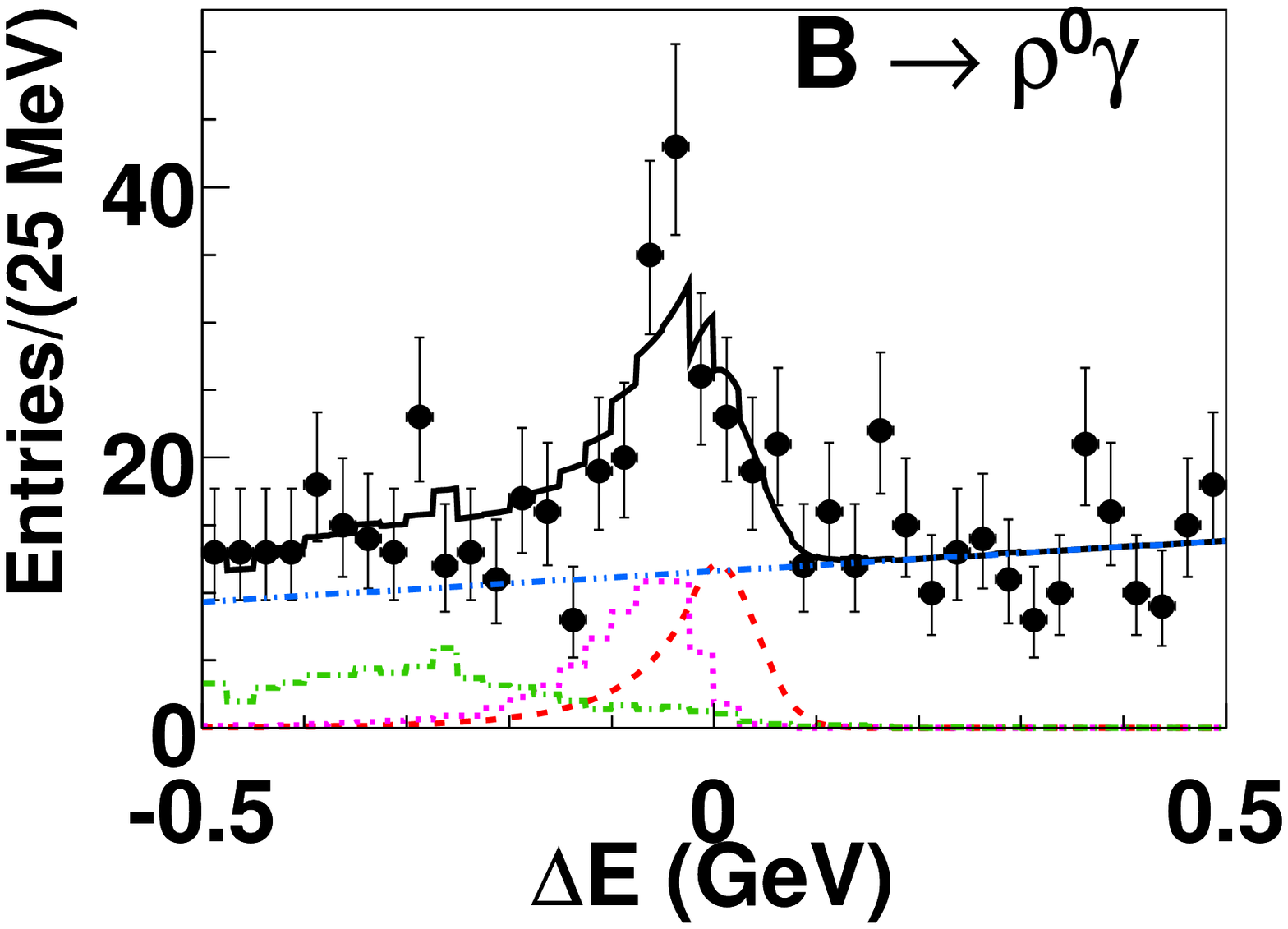}%
\vbox{
\hbox{\includegraphics[width=0.33\textwidth]{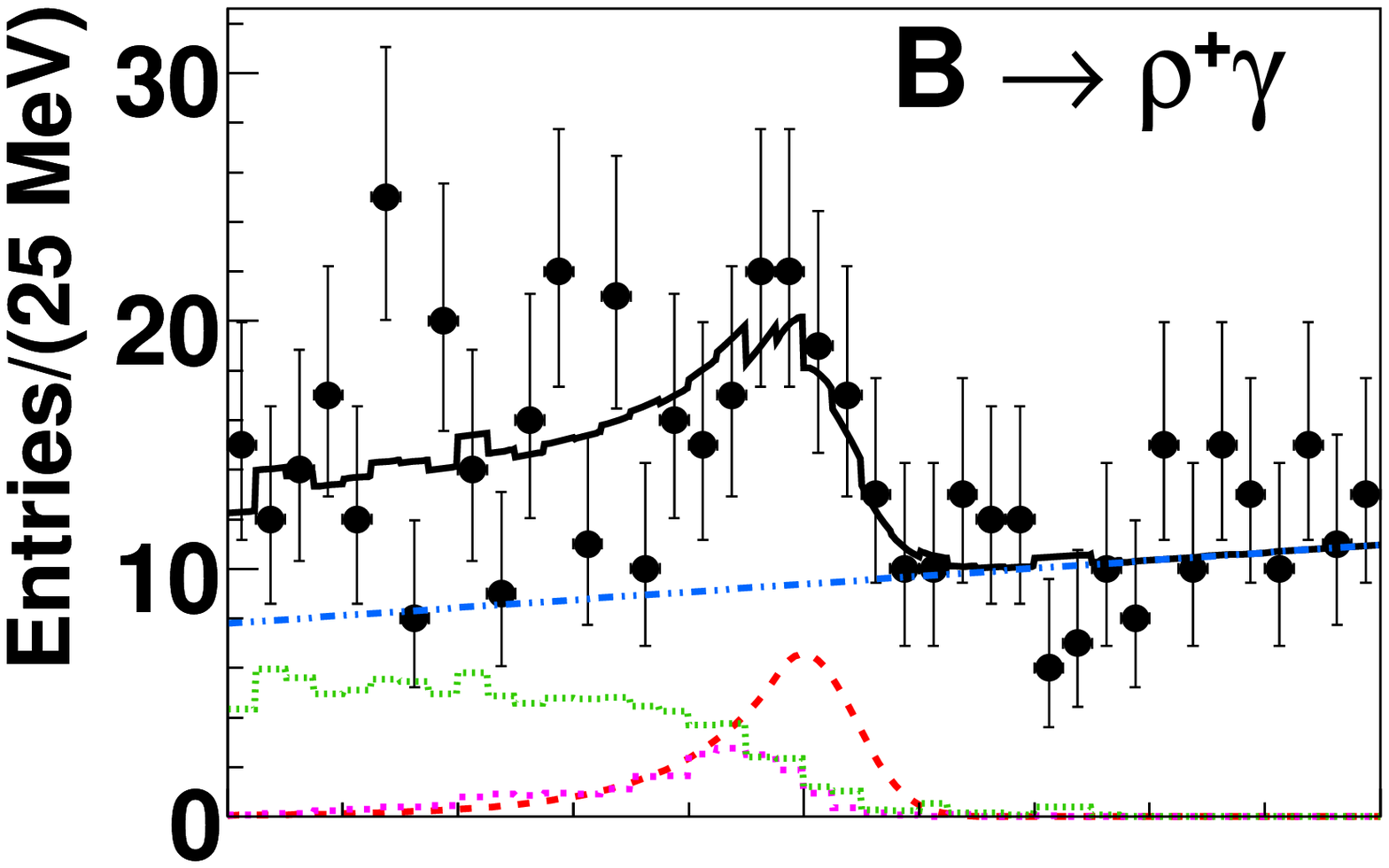}}\vspace{-0.065\textwidth}
\hbox to 0.33\textwidth{\hskip 0.044\textwidth
\includegraphics[width=0.28\textwidth]{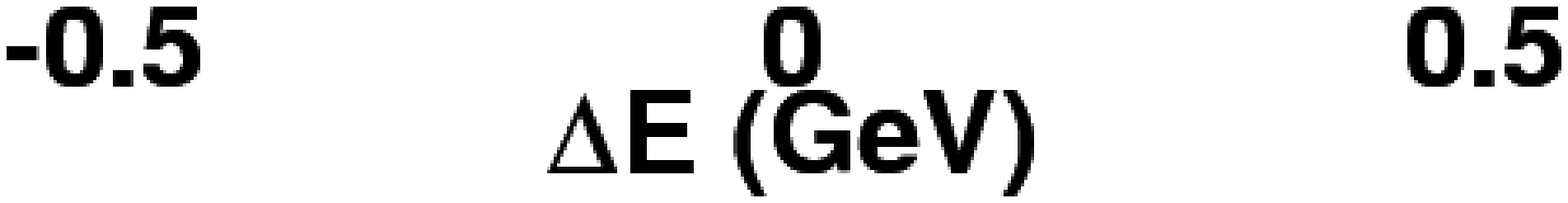}}\vspace{0.027\textwidth}
}%
\vbox{
\hbox{\includegraphics[width=0.33\textwidth]{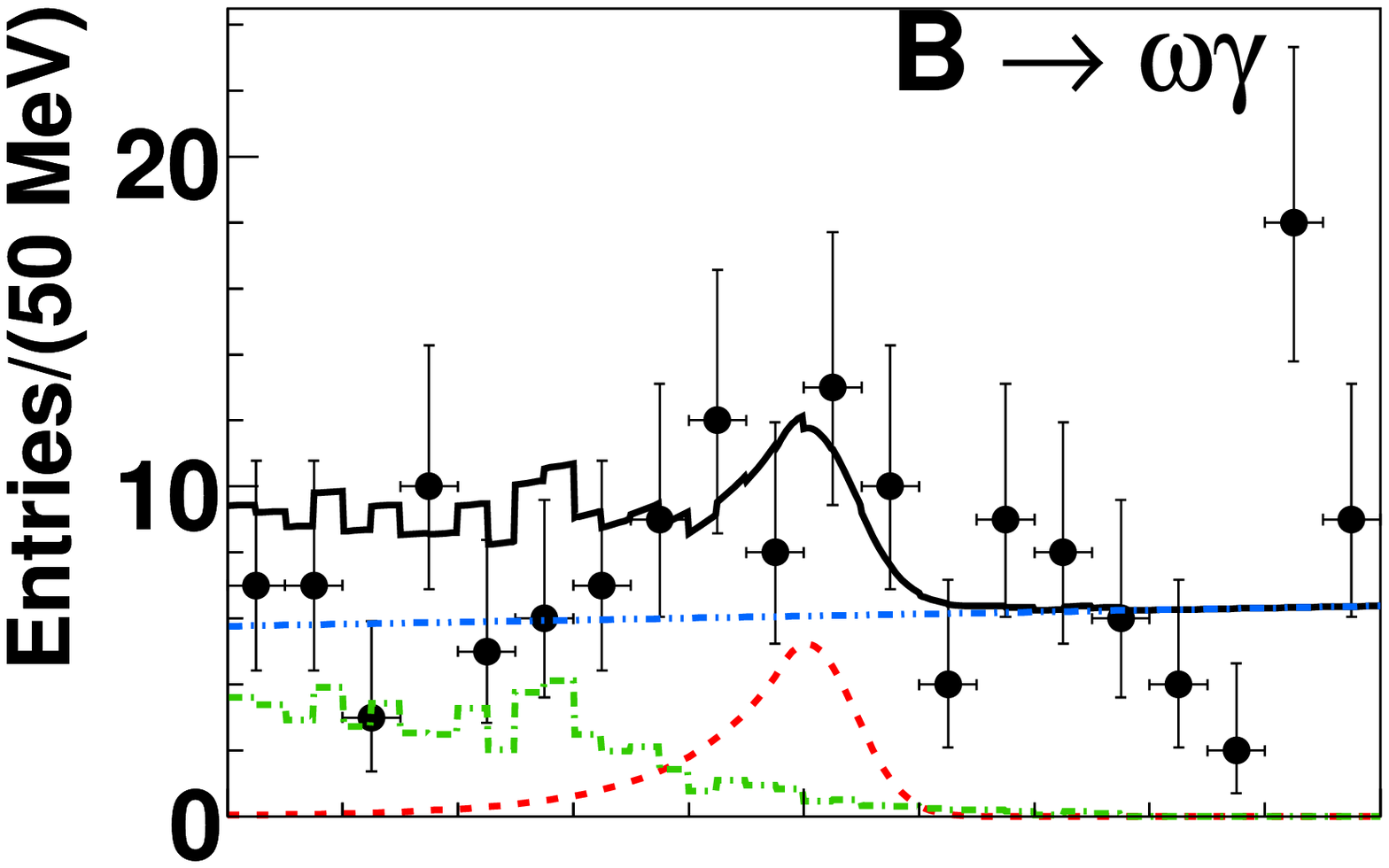}}\vspace{-0.065\textwidth}
\hbox to 0.33\textwidth{\hskip 0.044\textwidth
\includegraphics[width=0.28\textwidth]{fig/title_dE.eps}}\vspace{0.027\textwidth}
}
\vspace{-0.03\textwidth}
  \caption{Projections of the fit results to $\Delta E$ in
$M_\mathrm{bc}$ signal region for $B^+\to \rho^0\gamma$, $B^0\to
\rho^+\gamma$ and $B^0\to \omega\gamma$.  Curves show the signal
(dashed, red), continuum (dot-dot-dashed, blue), $B\to K^*\gamma$
(dotted, magenta), other backgrounds (dash-dotted, green), and the
total fit result (solid).
 \label{fig:btodg}}
\end{figure}

A fit to $M_\mathrm{bc}$ and $\Delta E$ (and $M_{K\pi}$ for the
$\rho^0\gamma$ mode) for candidates satisfying $|\Delta
E|<0.5\,\mathrm{GeV}$ and $M_\mathrm{bc}>5.2\,\mathrm{GeV}/c^2$
is performed individually for the three $\rho\gamma/\omega\gamma$
signal modes and the two control modes $K^+\gamma$ and $K^0\gamma$.
Significant $K^{*0}\gamma$, $K^{*+}\gamma$ background in the
$\rho^0\gamma$ and $\rho^+\gamma$ samples are shifted in $\Delta E$
from signal peak.  The shift offset of this background is determined
from $K^*$ enriched sample from data, which is also used to determine
the size using known kaon to pion misidentification rate.  The $\Delta E$
projections of fit results are shown in Fig.~ \ref{fig:btodg}.
Table~\ref{tbl:results} lists the obtained yields
and branching fractions~\cite{Taniguchi:2008ty}.

\def\BtoRG{B\to \rho\gamma}
\def\BtoRPG{B^+\to \rho^+\gamma}
\def\BtoRMG{B^-\to \rho^-\gamma}
\def\BtoRZG{B^0\to \rho^0\gamma}
\def\BtoROG{B\to (\rho,\omega)\gamma}
\def\BtoOMG{B^0\to \omega\gamma}
\def\BtoOG{B\to \omega\gamma}

\def\PM#1#2{\,^{+#1}_{-#2}{}}
\def\NSBtoRPG{45.8 \PM{15.2}{14.5} \PM{2.6}{3.9}}
\def\BrBtoRPG{8.7 \PM{2.9}{2.7} \PM{0.9}{1.1}} 

\def\NSBtoRZG{75.7 \PM{16.8}{16.0} \PM{5.1}{6.1}}
\def\BrBtoRZG{7.8 \PM{1.7}{1.6} \PM{0.9}{1.0}} 

\def\NSBtoOMG{17.5 \PM{8.2}{7.4} \PM{1.1}{1.0}}
\def\BrBtoOMG{4.0 \PM{1.9}{1.7}\pm 1.3} 

\def\signifRPconv{3.3}
\def\signifRZconv{5.0}
\def\signifOMconv{2.6}
\def\EffRP{8.03\pm0.59}
\def\EffRZ{14.81\pm0.95}
\def\EffOM{6.58\pm0.76}
\def\Br{{\cal B}}

\begin{table}
\caption{Yield, significance with systematic
  uncertainty, efficiency, and branching fraction ($\Br$) for each
  mode.  The first and second errors in the yield and $\Br$ are
  statistical and systematic, respectively. 
  The sub-decay
  $\Br(\omega\to\pi^+\pi^-\pi^0)$ is included for the $\omega\gamma$ mode.
}
\label{tbl:results}
\begin{tabular}{lcccc}\hline
Mode & Yield & Significance & Efficiency (\%) & $\Br$ ($10^{-7}$) \\
\hline
$\BtoRPG$ & $\NSBtoRPG$ & $\signifRPconv$ & $\EffRP$ & $\BrBtoRPG$ \\
$\BtoRZG$ & $\NSBtoRZG$ & $\signifRZconv$ & $\EffRZ$ & $\BrBtoRZG$ \\
$\BtoOMG$ & $\NSBtoOMG$ & $\signifOMconv$ & $\EffOM$ & $\BrBtoOMG$ \\
\hline
\end{tabular}
\end{table}

The branching fractions are combined to a single branching fraction
$\mathcal{B}(B\to (\rho,\omega)\gamma)\equiv \mathcal{B}(B^+\to
\rho^+\gamma) \equiv 2\times
\frac{\tau_{B^+}}{\tau_{B^0}}\mathcal{B}(B^0\to\rho^0\gamma) = 2\times
\frac{\tau_{B^+}}{\tau_{B^0}}\mathcal{B}(B^0\to\omega\gamma) $, and
the ratio to corresponding mode $B\to K^*\gamma$ is calculated as
$\frac{\mathcal{B}\to (\rho,\omega)\gamma)}{B\to K^*\gamma} =
0.0284\pm0.0050^{+0.0027}_{-0.0029}$.  This result is used to relate
the ratio of CKM matrix element: $|V_{td}/V_{ts}| =
0.195^{+0.020}_{-0.019}\mathrm{(exp.)}\pm 0.015\mathrm{(th.)}$.
The isospin asymmetry
$A_I(B\to\rho\gamma)=2{\tau_{B^+}\over\tau_{B^0}}
\mathcal{B}(B^0\to\rho^0\gamma)/\mathcal{B}(B^+\to\rho^+\gamma)-1$ is
calculated to be $A_I(B\to\rho\gamma)=
0.92^{+0.76}_{-0.71}{}^{+0.30}_{-0.35}$, which agrees with
BaBar~\cite{:2008gf} (note the different definition).  The direct
$CP$-violating asymmetry is also measured for the first time by
fitting $B^+\to\rho^+\gamma$ and $B^-\to\rho^-\gamma$ events
simultaneously, as
$A_{CP}(B^+\to\rho^+\gamma)=[N(\rho^-\gamma)-N(\rho^+\gamma)]
/[N(\rho^-\gamma)+N(\rho^+\gamma)] = {-0.11\pm{0.32}\pm{0.09}}$.




\begin{thebibliography}{9}   

\bibitem{Abe:2008sx} K.~Abe {\it et al.}  [Belle Collaboration],
  arXiv:0804.1580 [hep-ex].  

\bibitem{Li:2008qm}
  J.~Li {\it et al.}  [Belle Collaboration],
  arXiv:0806.1980 [hep-ex].

\bibitem{Atwood:2004jj}
  D.~Atwood, T.~Gershon, M.~Hazumi and A.~Soni,
  Phys.\ Rev.\  D {\bf 71}, 076003 (2005)
  [arXiv:hep-ph/0410036].


\bibitem{Taniguchi:2008ty}
  N.~Taniguchi {\it et al.}  [Belle Collaboration],
  Phys.\ Rev.\ Lett.\  {\bf 111}, 111801 (2008)
  [arXiv:0804.4770 [hep-ex]].

\bibitem{:2008gf}
B.~Aubert {\it et al.}  [BABAR Collaboration],
  arXiv:0808.1379 [hep-ex].

\end{thebibliography}
\end{document}